\documentclass[prl,showpacs,amsmath,amssymb,twocolumn, 9pt]{revtex4}

\usepackage{amsthm}
\usepackage{dcolumn}
\usepackage{bm}
\usepackage{graphicx}

\begin{document}

\newtheorem{corollary}{Corollary}
\newtheorem{definition}{Definition}
\newtheorem{example}{Example}
\newtheorem{lemma}{Lemma}
\newtheorem{proposition}{Proposition}
\newtheorem{theorem}{Theorem}
\newtheorem{fact}{Fact}
\newtheorem{property}{Property}
\newcommand{\bra}[1]{\langle #1|}
\newcommand{\ket}[1]{|#1\rangle}
\newcommand{\braket}[3]{\langle #1|#2|#3\rangle}
\newcommand{\ip}[2]{\langle #1|#2\rangle}
\newcommand{\op}[2]{|#1\rangle \langle #2|}

\newcommand{\tr}{{\rm tr}}
\newcommand {\E } {{\mathcal{E}}}
\newcommand {\F } {{\mathcal{F}}}
\newcommand {\diag } {{\rm diag}}
\newcommand{\slocc}{\overset{\underset{\mathrm{SLOCC}}{}}{\longrightarrow}}

\title{\Large {\bf Obtain $W$-state from $GHZ$-state with asymptotic rate $1$}}
\author{Nengkun Yu$^{1,2}$}
\email{nengkunyu@gmail.com}
\author{Cheng Guo$^{2}$}
\author{Runyao Duan$^{2,1,3}$}
\affiliation{$^1$State Key Laboratory of Intelligent Technology and Systems,Tsinghua National Laboratory for Information Science and Technology, Department of Computer Science and Technology, Tsinghua University, Beijing 100084, China}
\affiliation{$^2$Centre for Quantum Computation and Intelligent Systems (QCIS), Faculty of Engineering and Information Technology, University of Technology, Sydney (UTS), NSW 2007, Australia}
\affiliation{$^3$UTS-AMSS Joint Research Laboratory for Quantum Computation and Quantum Information Processing, Academy of Mathematics and Systems Science, Chinese Academy of Sciences, Beijing 100190, China}
\begin{abstract}
We introduce a notion of entanglement transformation rate to characterize the asymptotic comparability of two multipartite pure entangled states under stochastic local operations and classical communication (SLOCC). For two well known SLOCC inequivalent three-qubit states $\ket{GHZ}=\tfrac{1}{\sqrt{2}}(\ket{000}+\ket{111})$  and $\ket{W}=\tfrac{1}{\sqrt{3}}(\ket{100}+\ket{010}+\ket{001})$, we show that  the entanglement  transformation rate from $\ket{GHZ}$ to $\ket{W}$ is exactly $1$. That means that we can obtain  one copy of $W$-state, from one copy of $GHZ$-state by SLOCC, asymptotically. We then apply similar techniques to obtain a lower bound on the entanglement transformation rates from an $N$-partite $GHZ$-state to a class of Dicke states, and prove the tightness of this bound for some special cases that naturally generalize the $\ket{W}$ state. A new lower bound on the tensor rank of matrix permanent is also obtained by evaluating the the tensor rank of Dicke states.
\end{abstract}

\pacs{03.65.Ud, 03.67.Hk}

\maketitle
\textit{Introduction---} Multipartite entanglement has been widely studied \cite{BBPS96,BPR+01,BC01,KEM99} since it is a proven asset to information
processing and computational tasks. In order to quantitatively compare between different types of quantum information resources, the following fundamental entanglement transformation problem arises: whether a pure $N$-partite state $\ket{\psi}$ can be transformed into another given $N$-partite state $\ket{\phi}$, assuming that each party may perform only local operations on their respective systems with the help of unlimited two-way classical communication(LOCC). In bipartite case, a necessary and sufficient condition for entanglement transformation was reported by Nielsen in \cite{NIE99}. His result indicates that the vector of Schmidt coefficients, instead of any scalar, is a proper entanglement measure when exact transformations are considered. After that, multiple-copy entanglement transformation was studied: Duan \textit{et al} proved that entanglement-assisted transformation of bipartite case is asymptotically equivalent to multiple-copy transformation \cite{DFY05}; in a multipartite setting, Ji \textit{et al} showed that the entanglement transformation rate between any two genuinely entangled states are positive, that is, it is always feasible to exactly transform a genuinely $N$-partite entangled pure state with sufficient many  but a finite number of copies to any other $N$-partite state by LOCC \cite{JDY04}, where a multipartite pure entangled state is said to be \textit{genuinely} if it is not in a product form between any bipartite partition of the parties.

One of the major difficulties in evaluating the entanglement transformation rate of multipartite case is that the class of LOCC is still not satisfactorily understood. Another one is the richness of multipartite entanglement. Generally, there exist incomparable states, even in three-qubit systems. It is still unclear how to determine whether  one multipartite state can be transformed to another by LOCC. To partially remedy these obstacles, we relax the restriction of LOCC and consider the class of stochastic local operations and classical communication (SLOCC) \cite{EB01,HEB04,VDMV02,YU13,DVC00}. The ability to transform a state $\ket{\psi}$ to another state $\ket{\phi}$ with SLOCC is symbolically expressed as $\ket{\psi}\slocc\ket{\phi}$. The physical meaning of SLOCC operations is that they can be implemented by LOCC operations with nonzero probability. In fact, SLOCC has been used to study entanglement classification \cite{DVC00,GW13} and entanglement transformation \cite{CDS08,YCGD10,CCDJ+10}. The whole multipartite state space can be divided into SLOCC equivalence classes. For instance, D$\mathrm{\ddot{u}}$r \textit{et al\!} observed that within three qubit systems, there exist two distinct equivalence classes of genuinely tripartite entangled states,$\ket{GHZ}=\tfrac{1}{\sqrt{2}}(\ket{000}+\ket{111})$ and $\ket{W}=\tfrac{1}{\sqrt{3}}(\ket{100}+\ket{010}+\ket{001})$ \cite{DVC00}.

Comparing with LOCC entanglement transformation, SLOCC entanglement transformation of pure states has a much simpler mathematical structure that can be directly characterized. In order to consider the asymptotic SLOCC entanglement transformation between pure states, we only need to deal with SLOCC equivalent classes.  By employing the concept of tensor rank, which is defined as the smallest number of product states whose linear span contains the given state, many interesting results are obtained. For three-qubit systems, it was showed that $3$ copies of $GHZ$-state can be transformed into $2$ copies of $W$-state \cite{CDS08}. In \cite{YCGD10}, we proved that $\ket{GHZ}^{\otimes m}\slocc\ket{W}^{\otimes 2n}$ is valid if $2^m\geq 7^n$. Later, it was demonstrated that SLOCC protocol can transform $4$ copies of $GHZ$-state to $3$ copies of $W$-state \cite{CCDJ+10}. These increasing lower bounds reflect both the richness of entanglement and the difficulty of obtaining asymptotic results. These progresses motivate us to introduce a useful  notion of the \textit{SLOCC entanglement transformation rate} in the following way:
\begin{equation*}
R(\ket{\psi},\ket{\phi})=\mathrm{sup}\{\frac{m}{n}:\ket{\psi}^{\otimes n}\slocc\ket{\phi}^{\otimes m}\}.
\end{equation*}
This quantity intuitively characterizes the optimal number of copies of $\ket{\phi}$ one can obtain from a single copy of $\ket{\psi}$ under SLOCC, in an asymptotic setting. Therefore, it  is of great interest to determine this value by studying the highest possible rate for the multi-copy transformations. Unfortunately, $R(\ket{\psi},\ket{\phi})$ is not easy to calculate, even for the simplest non-trivial case, $R(\ket{GHZ},\ket{W})$, whose exact value was conjectured to be $1$ \cite{YCGD10,CCDJ+10}.

In this Letter, we prove the validity of the above conjecture by constructing an SLOCC transformation from $n+o(n)$ copies of $GHZ$-state to $n$ copies of $W$-state, that is, one can obtain $1$ copy
of $W$-state, from $1$ copy of $GHZ$-state by SLOCC, asymptotically. To reach our goal, we introduce a class of tripartite
states such that $n$ copies of $W$-state can be written into the sum of at most $n^2$ items of such states,
and each state of this class can be obtained by applying SLOCC operations on $n$ copies of $GHZ$-state. Then,
we show that $R(\ket{GHZ}_N,\ket{W}_N)$, the entanglement transformation rate, is also $1$ for $N$-partite state
$\ket{GHZ}_N=\tfrac{1}{\sqrt{2}}(\ket{00\cdots 0}+\ket{11\cdots 1})$ and $\ket{W}_N=\frac{1}{\sqrt{N}}(\ket{0\cdots 01}+\cdots+\ket{10\cdots 0})$.
We generalize this technique to obtain a lower bound on the entanglement transformation rate that from a general $GHZ$-state to a class of
fully symmetric states, Dicke states. More precisely,
we show that $(\sum_{i=2}^d \log_{2}(j_i+1))^{-1}$ is a lower bound of $R(\ket{GHZ}_N,\ket{D(j_1,\cdots,j_d)})$.
Here the Dicke state $\ket{D(j_1,\cdots,j_d)}$ is defined as follows,
\begin{equation}\label{Dicke}
  \ket{D(j_1,\cdots,j_d)} \! := \! {N \choose j_1 \ldots j_d}^{\!-1/2} \!\!
               P_{\textrm{sym}}
               \bigl(\otimes_{i=1}^d \ket{i}^{\otimes j_i} \bigr),
\end{equation}
where $\{ \ket{1},\ldots,\ket{d} \}$ is a computational
basis of the $d$-dimensional Hilbert space $\mathcal{H}_d$, $P_{\textrm{sym}}$ is the projection onto the Bosonic (fully symmetric)
subspace, $i.e.$, $P_{\textrm{sym}} = \frac{1}{N!} \sum_{\pi\in S_N} U_\pi$, the
sum extending over all permutation operators $U_\pi$ of the $N$ systems, $N=\sum_{i=1}^d j_i$, and $j_1\geq \cdots \geq j_d\geq 1$. For the special case such as $j_1\geq \sum_{i=2}^d j_i$, the tightness of this bound is proved. At last, we study the tensor rank
of permanent of a matrix, one of the most
extensively studied computational problems, by simply evaluating the tensor rank of the Dicke state
$\ket{D(1,\cdots,1)}$.

\textit{Main Results---}Our first result is the following,
\begin{theorem}\label{theorem1}
For three-qubit state system, we have
$$R(\ket{GHZ},\ket{W})=1.$$
That is, for sufficient large $n$, one can transform $n+o(n)$ copies of $GHZ$-state to $n$ copies of $W$-state by SLOCC. An immediate consequence is that the $GHZ$ state is asymptotically stronger than the $W$ state under SLOCC, although they are incomparable at the single copy level.

Generally, for $N$-partite systems,
$$R(\ket{GHZ}_N,\ket{W}_N)=1.$$
Again one can obtain $\ket{W}_N$ from $\ket{GHZ}_N$ at a rate $1$ by SLOCC.
\end{theorem}

For convenience,  in the following discussions we omit an unimportant normalized factor and denote directly $\ket{W}=\ket{100}+\ket{010}+\ket{001}$. Before proving the validity of Theorem \ref{theorem1} for three-qubit systems, we first introduce a class of tripartite states, $\ket{[a,b,c]}_n$ for any triple $(a,b,c)$ of nonnegative integers such that $a+b+c=n$. Let $B$ be the following set
\begin{equation*}
B:=\{(a,b,c):a+b+c=n,0\leq a,b,c\leq n\}.
\end{equation*}
For any $(a,b,c)\in B$, one can define an unnormalized state
\begin{equation*}
\ket{[a,b,c]}_n=\sum_{ \substack{i\oplus j\oplus k=(11\cdots 1)_n,\\ i\in A(a),j\in A(b),k\in A(c)}}\ket{i}\ket{j}\ket{k},
\end{equation*}
where $\oplus$ is the bitwise addition modulo 2, $(11\cdots 1)_n$ stands for the $n-$bit string with `1' in all $n$ positions, and $A(\cdot)$ represents the set of the $n-$bit strings with the same Hamming weight, i.e.,
\begin{equation*}
A(l)=\{i:h(i)=l,i\in \mathbb{Z}_2^n\},
\end{equation*}
where $\mathbb{Z}_2=\{0,1\}$ and the Hamming weight $h(i)$ of an $n$-bit string $i$ simply represents the number of `1's in the string.

We can verify the following equation,
\begin{equation}\label{Wn}
\ket{W}^{\otimes n}=\sum_{(i,j,k)\in S}\ket{i}\ket{j}\ket{k}=\sum_{(a,b,c)\in B}\ket{[a,b,c]}_n,
\end{equation}
where $i,j,k$ are $n$-bit strings, and $S$ is the subset of $\mathbb{Z}^n_2\times\mathbb{Z}^n_2\times\mathbb{Z}^n_2$,
\begin{equation*}
S=\{(i,j,k):i\oplus j\oplus k=(11\cdots 1)_n, i_tj_tk_t=0~\rm{for~any~} t\},
\end{equation*}
with $i_t$ the $t^{th}$ bit of $i$.
Namely, $S$ is the set of $(i,j,k)$ such that on each $1\leq t\leq n$, there is only a single `1' in the $t$-th bit of $i$, $j$, and $k$.
The first equality in Eq. (\ref{Wn}) follows by calculating $\ket{W}^{\otimes n}$ in computational basis, while the second equality follows by observing that

\begin{equation*}
S=\{(i,j,k):i\oplus j\oplus k=(11\cdots 1)_n, h(i)+h(j)+h(k)=n\}.
\end{equation*}

The following property of tripartite states $\ket{[a,b,c]}_n$ is extremely useful in proving Theorem 1,
\begin{lemma}
Any state $\ket{[a,b,c]}_n$ can be obtained from $\ket{GHZ}^{\otimes n}$ by SLOCC.
\end{lemma}
\textit{Proof:---} To see the validity of this lemma, we need the following identity,
\begin{widetext}
\begin{equation*}
\ket{[a,b,c]}_n=\frac{1}{2^n}\sum_{l=0}^{2^n-1}(\sum_{i\in A(a)}(-1)^{l\cdot \overline{i}}\ket{i})\otimes(\sum_{j\in A(b)}(-1)^{l\cdot \overline{j}}\ket{j})\otimes(\sum_{k\in A(c)}(-1)^{l\cdot \overline{k}}\ket{k}).
\end{equation*}
\end{widetext}
Here $\overline{i}=(11\cdots 1)_n\oplus (i_1i_2\cdots i_n)$, and $l\cdot \overline{i}$ is the bitwise inner product of two $n-$bit strings $l$ and $\overline{i}$. The above identity can be verified by a direct calculation.

We construct three $2^n\times 2^n$ matrices
\begin{eqnarray*}
E=\sum_{l=0}^{2^n-1}\sum_{i\in A(a)}(-1)^{l\cdot \overline{i}}\op{i}{l},\\
F=\sum_{l=0}^{2^n-1}\sum_{j\in A(b)}(-1)^{l\cdot \overline{j}}\op{j}{l},\\
G=\sum_{l=0}^{2^n-1}\sum_{k\in A(c)}(-1)^{l\cdot \overline{k}}\op{k}{l},
\end{eqnarray*}
Now, it is direct to verify that
$$(E\otimes F\otimes G)\ket{GHZ}^{\otimes n}={2^{n/2}}\ket{[a,b,c]}_n,$$
where we have assumed that $$\ket{GHZ}^{\otimes n}=\frac{1}{2^{n/2}}\sum_{l=0}^{2^{n}-1}\ket{l}\ket{l}\ket{l}.$$ That is, $E\otimes F\otimes G$ transforms
$\ket{GHZ}^{\otimes n}$ to $\ket{[a,b,c]}_n$.
\hfill $\blacksquare$

Back to the proof of Theorem \ref{theorem1},

\textit{Proof of Theorem 1:---} Let $|B|$ be the cardinality of $B$. Noticing that $$|B|=\left(\begin{array}{cc}n+2\\
2\end{array}\right)=O(n^2),$$ we conclude that
\begin{equation*}
\ket{GHZ}^{\otimes m}\slocc\ket{W}^{\otimes n},
\end{equation*}
holds for all $2^m\geq \left(\begin{array}{cc}n+2\\
2\end{array}\right)2^{n}$.
Therefore,
$$R(\ket{GHZ},\ket{W})=1.$$
This method can also be used to show that for any $N>1$,
\begin{equation*}
\ket{GHZ}_N^{\otimes m}\slocc\ket{W}_N^{\otimes n},
\end{equation*}
holds when $2^m\geq \left(\begin{array}{cc}n+N-1\\
N-1\end{array}\right)2^{n}$.

This lead us to the fact that
$$R(\ket{GHZ}_N,\ket{W}_N)\geq 1.$$
On the other hand, it is known that the tensor rank of $\ket{W}_N^{\otimes n}$ is no less than $(N-1)2^n-N+2$ \cite{CCDJ+10}, which impies
$$R(\ket{GHZ}_N,\ket{W}_N)\leq 1.$$
Combining with these results, we know that $R(\ket{GHZ}_N,\ket{W}_N)=1.$ \hfill $\blacksquare$

Both $\ket{W}$ and $\ket{GHZ}$ are symmetric states, i.e., those invariant under any permutation of its parties. Dicke states, defined in Eq. (\ref{Dicke}), are a natural generalization of these states. Studying the entanglement measure of such states has attracted a lot of attention\cite{HKWG+09,TG09}. The entanglement transformation properties of such states are studied \cite{BKMG+09,CCDJ+10} and some families of multi-qubit SLOCC
equivalent states are realized by using symmetric states \cite{BTZLS+09,WKSW09}.
In order to generalize Theorem 1 to Dicke states, we need evaluate the entanglement transformation rate of Dicke states by SLOCC. A general lower bound is given as follows.
\begin{theorem}
For Dicke state $\ket{D(j_1,\cdots,j_d)}$ with $j_1\geq \cdots\geq j_d$ and $N=\sum_{i=1}^d j_i$,
\begin{eqnarray*}
R(\ket{GHZ}_N,\ket{D(j_1,\cdots,j_d)})\geq (\sum_{i=2}^d\log_{2}{(j_i+1)})^{-1}.
\end{eqnarray*}
The bound is tight for Dicke state with $j_1\geq \sum_{i=2}^d j_i$.
\end{theorem}

\textit{Proof:---}The proof of the lower bound part is the direct generalization of Theorem \ref{theorem1}. For readability, we postpone the detailed proof of
this part to the Supplementary Materials \cite{SM}.

In order to show the tightness of the above bound for Dicke state with $j_1\geq \Pi_{i=2}^d(j_i+1)$, we regard $\ket{D(j_1,\cdots,j_d)}$ as a bipartite state $\ket{\psi}$ by arranging the first $r$ parties of $\ket{D(j_1,\cdots,j_d)}$ into a single party, say Alice, and the rest $N-r$ parties into another single party, Bob, where $r=\lfloor N/2\rfloor$. From the definition, we know that the tensor rank of $\ket{D(j_1,\cdots,j_d)}$ is not less than that of $\ket{\psi}$. Now, we apply local operator $M\otimes N$ on $\ket{\psi}$, where $M$ maps $\ket{\alpha_1}\ket{\alpha_2}\cdots\ket{\alpha_r}$ into $\ket{1}^{\otimes \mu_1}\ket{2}^{\otimes \mu_2}\cdots\ket{d}^{\otimes \mu_d}$, where $\mu_i$ are the multiplicity of $i$ among $\alpha_1,\alpha_2,\cdots,\alpha_r$. The definition of $N$ is similar.

Observe that the tensor rank of $(M_A\otimes N_B)\ket{\psi}$ equals to $f$ which is the cardinality of the following set,
$$\{(\beta_1,\beta_2,\cdots,\beta_d):0\leq\beta_i\leq j_i,\sum_{i=1}^d\beta_i=r\}.$$
For Dicke state with $j_1\geq \sum_{i=2}^d j_i$, we observe that $j_1\geq N/2\geq r$. Thus the constraint $0\leq \beta_1\leq j_1$ in the above set is automatically satisfied and the cardinality is totally determined by other $\beta_i$ such that $i\geq 2$. By a simple counting arguments we know the cardinality is given by $\Pi_{i=2}^d(j_i+1)$, which is also the tensor rank of the bipartite state.
Noticing that the tensor rank of bipartite pure state is multiplicative, and tensor rank is a
strictly non-increasing quantity under SLOCC, we conclude that
$$R(\ket{GHZ}_N,\ket{D(j_1,\cdots,j_d)})\leq (\sum_{i=2}^d\log_{2}{(j_i+1)})^{-1},$$
for $j_1\geq \Pi_{i=2}^d(j_i+1)$.
Thus, $(\sum_{i=2}^d\log_{2}{(j_i+1)})^{-1}$ is the tight bound for such Dicke state.
\hfill $\blacksquare$

By applying the above technique to the $N-$partite state $\ket{D(1,\cdots,1)}$, we have the following result about the SLOCC transformation,
\begin{lemma}\label{lemma2}
The tensor rank of $ \ket{D(1,\cdots,1)}$ is not less than $\left(\begin{array}{cc}N\\
\lfloor N/2\rfloor\end{array}\right)$. Thus,
\begin{equation*}
\ket{GHZ}_N^{\otimes m}\slocc \ket{D(1,\cdots,1)} \Longrightarrow 2^m\geq \left(\begin{array}{cc}N\\
\lfloor N/2\rfloor\end{array}\right),
\end{equation*}
\end{lemma}

Together with the known result that the tensor rank of $\ket{D(1,\cdots,1)}$ is not more than $\leq 2^{N-1}$ from \cite{CCG12}, we can conclude that its tensor rank is $2^{N(1+o(1))}$.

The motivation for studying the tensor rank of $\ket{D(1,\cdots,1)}$ is the connection between its tensor rank and that of matrix permanent, a homogeneous polynomial.
It is worth noting that tensor rank of homogeneous polynomial has already been extensively studied in algebraic complexity theory \cite{BCS97,HAA90,CGL08,LT10,RAZ04,RY09,RAZ10}. A homogeneous polynomial is a multi-variables polynomial whose nonzero terms (monomials) all have the same degree.The tensor rank of a homogeneous polynomial $P(x_1,\cdots,x_n)$ is defined as the smallest number of $rk$ such that $P(x_1,...x_n)$ can be written as the sum of $rk$ terms of $L^{i}(x_1,\cdots, x_n)$, where each $L^{i}(\cdots)$ is the product of $d$ homogenous linear forms with $d$ the degree of the polynomial.

The permanent of matrix $X = (x_{i,j})_{N\times N}$ is defined as
\begin{eqnarray*}
    \operatorname{perm}(X)=\sum_{\sigma\in S_n}\prod_{i=1}^N x_{i,\sigma(i)}.
\end{eqnarray*}
The sum here extends over all elements $\sigma$ of the symmetric group $S_N$; $i.e.$ over all permutations of the numbers $1, 2,\cdots, N$.

It is easy to see that the tensor rank of matrix permanent is defined as the minimum number $rk$ such that
\begin{eqnarray*}
    \operatorname{perm}(X)=\sum_{j=1}^{rk}\prod_{i=1}^N L^{i,j}(X),
\end{eqnarray*}
where $L^{i,j}(X)$ is a linear function of $X$ and $L^{i,j}(0)=0$.

The tensor rank of matrix permanent is still unknown, and it relates to the central problem of computational complexity theory--circuit lower bounds.
One possible direction to study this problem is to restrict the form of $\prod_{i=1}^N L^{i,j}(X)$,
for instance, assume that $\prod_{i=1}^N L^{i,j}(X)$ satisfy multilinear property, see \cite{RAZ04,RAZ10,RY09}. In this case, it is known that the tensor rank of matrix permanent is lower bounded by $2^{N^{\Omega(1)}}$, where $\Omega(1)$ is some non-zero constant.

For a more restricted form, each $L^{i,j}(X)$ only depends on the $i$-th row of $X$ and $j$, we can obtain a better lower bound as follows.
\begin{theorem}
Let $X$ be an $N\times N$ matrix with permanent
\begin{eqnarray*}
    \operatorname{perm}(X)=\sum_{j=1}^{k(N)}\prod_{i=1}^N \sum_{k=1}^N a_{i,k}^{(j)}x_{i,k}.
\end{eqnarray*}
Then $k(N)\geq \left(\begin{array}{cc}N\\
\lfloor N/2\rfloor\end{array}\right)$.
\end{theorem}
\textit{Proof:---} Indeed, since each $L^{i,j}(X)$ only depends on the $i$-th row of $X$ and $j$, we can easily verify that
$$\ket{D(1,\cdots,1)}=\sum_{j=1}^{k(N)}\prod_{i=1}^N (\sum_{k=1}^N a_{i,k}^{(j)}\ket{k}).$$
The proof is completed by applying Lemma \ref{lemma2}.
\hfill $\blacksquare$

\textit{Conclusions---}
There are still many unsolved problems concerning the SLOCC transformation and tensor rank. For instance, it would be of great interest to obtain the precise value of the entanglement transformation rate from an $N$-partite $GHZ$-state to a general Dicke state $\ket{D(j_1,\cdots, j_d)}$. In particular, it seems quite intriguing to determine whether the lower bound $(\sum_{i=2}^d \log_2 (j_i+1))^{-1}$ is actually tight in general. That would require new techniques to lower bound the tensor rank of muli-copy Dicke states. We also hope our work will help to study homogeneous polynomials from geometric perspective, for instance, introducing new ideas and techniques from quantum information theory to algebraic complexity theory.

We thank Xiaoming Sun, Youming Qiao, and Mingsheng Ying for helpful discussions and comments. This work was partly supported by the National Natural Science Foundation of China (Grant No. 61179030) and the Australian Research Council Discovery Projects (Grant Nos. DP110103473 and DP120103776). R.D. was also supported in part by an ARC Future Fellowship (Grant No. FT120100449).

\textit{Proof of the first part of Theorem 2:---}
We will prove that
\begin{equation*}
\ket{GHZ}_N^{\otimes m}\slocc\ket{D(j_1,\cdots,j_d)}^{\otimes n}
\end{equation*}
holds if
$$2^m\geq \prod \limits _{k=2}^{d}\left(\begin{array}{cc}nj_k+N-1\\
N-1\end{array}\right)(j_k+1)^n.$$
The proof is a nontrivial generalization of the $\ket{W}$ state case.

Before presenting the proof, we introduce some notations: $T$ is used to denote $\{\ket{1},\cdots,\ket{d}\}^{\otimes n}$, where the tensor product $\mathrm{S}_1\otimes \mathrm{S}_2$ of two sets $\mathrm{S}_1$ and $\mathrm{S}_2$ is defined as $\{\ket{s_1}\otimes{\ket{s_2}:\ket{s_k}\in \mathrm{S}_k}, k=1,2\}$.
Now we associate any $\ket{\alpha}\in T$ with a $d-1$ dimensional vector of natural numbers $\vec{v}=(l_2,\cdots,l_d)$ iff the number of $\ket{k}$ appearing in $\ket{\alpha}$ is $l_k$ for all $2\leq k\leq d$, $\vec{v}$ is called the characteristic vector of $\ket{\alpha}$, written as $C(\ket{\alpha})=\vec{v}$.

Now we divide $T$ into disjoint subsets $A_{\vec{v}}$ according to their characteristic vector:
$A_{\vec{v}}=\{\ket{\alpha}:C(\ket{\alpha})=\vec{v}.\}$

Define $B$ as the following set
\begin{equation*}
B=\{(\vec{v}_1,\cdots,\vec{v}_d):\vec{v}_1+\cdots+\vec{v}_d=(nj_2,\cdots,nj_d)\},
\end{equation*}
where $\vec{l}_k$ are all vectors of natural numbers.

Now we decompose $\ket{D(j_1,\cdots,j_d)}^{\otimes n}$ according to the computational basis, and rearrange the elements according to characteristic vectors as follows
\begin{equation*}
\ket{D(j_1,\cdots,j_d)}^{\otimes n}=\sum_{(\vec{v}_1,\cdots,\vec{v}_d)\in B}\ket{[\vec{v}_1,\cdots,\vec{v}_d]}_n.
\end{equation*}
Note that here $\ket{[\vec{v}_1,\cdots,\vec{v}_d]}_n$ is not the superposition of all $\ket{\alpha_1}\otimes\cdots\otimes\ket{\alpha_d}$ with $C(\ket{\alpha_k})=\vec{v}_k$. We also require that $\ket{\alpha_1}\otimes\cdots\otimes\ket{\alpha_d}$ does appear in the decomposition of $\ket{D(j_1,\cdots,j_d)}^{\otimes n}$.

Noticing that $|B|$ is a polynomial in $n$, we only need to show
for any $(\vec{v}_1,\cdots,\vec{v}_d)\in B$ with $2^m \geq \prod \limits _{k=2}^{d}(j_k+1)^n$
$$\ket{GHZ}_N^{\otimes m}\slocc\ket{[\vec{v}_1,\cdots,\vec{v}_d]}_n.$$
This is proved by verifying the following equation with $w_t:=e^{\frac{2\pi i}{t}}$, $\mu(L):=\prod\limits_{k=2}^d w_{j_k+1}^{\sum_{p=1}^n l_{k,p}}$, and
$f(\alpha,L):=\prod\limits_{p=1}^n w_{j_{s_p}+1}^{l_{s_p,p}}$ for $\ket{\alpha}=\ket{s_1}\otimes\cdots\otimes\ket{s_n}$ with $s_p\in \{1,\cdots d\}$,
\begin{widetext}
\begin{equation*}
\ket{[\vec{v}_1,\cdots,\vec{v}_d]}_n=\frac{1}{\prod \limits _{k=2}^{d}(j_k+1)^n}\sum_{\substack{l_{d,1}=0\\\cdots\\ l_{d,n}=0}}^{j_d}\cdots\sum_{\substack{l_{2,1}=0\\\cdots\\ l_{2,n}=0}}^{j_2}\mu(L)(\sum_{C(\ket{\alpha_1})=\vec{v}_1}f(\alpha_1,L)\ket{\alpha_1})\otimes\cdots\otimes(\sum_{C(\ket{\alpha_d})=\vec{v}_d}f(\alpha_d,L)\ket{\alpha_d}).
\end{equation*}
\end{widetext}
\end{document}